\documentclass[a4paper]{article}

\usepackage{INTERSPEECH2021}
\usepackage{graphicx}
\usepackage{subfigure}
\usepackage{array}
\usepackage{float}
\usepackage{algorithm}
\usepackage{enumitem}
\usepackage{multirow}
\usepackage{diagbox}
\usepackage{color}

\title{Separation Guided Speaker Diarization in Realistic Mismatched Conditions}
\name{Shu-Tong Niu$^1$, Jun Du$^1$, Lei Sun$^2$,  Chin-Hui Lee$^3$}
\address{
	$^1$University of Science and Technology of China, Hefei, Anhui, P.R.China\\
	$^2$iFlytek Research, Hefei, Anhui, P. R. China\\
	$^3$Georgia Institute of Technology, Atlanta, GA, USA}
\email{niust@mail.ustc.edu.cn, jundu@ustc.edu.cn, sunlei17@mail.ustc.edu.cn, chl@ece.gatech.edu}

\begin{document}

\maketitle
\begin{abstract}
We propose a separation guided speaker diarization (SGSD) approach by fully utilizing a complementarity of speech separation and speaker clustering. Since the conventional clustering-based speaker diarization (CSD) approach cannot well handle overlapping speech segments, we investigate, in this study, separation-based speaker diarization (SSD) which inherently has the potential to handle the speaker overlap regions. Our preliminary analysis shows that the state-of-the-art Conv-TasNet based speech separation, which works quite well on the simulation data, is unstable in realistic conversational speech due to the high mismatch speaking styles in simulated training speech and read speech. In doing so, separation-based processing can assist CSD in handling the overlapping speech segments under the realistic mismatched conditions.  Specifically, several strategies are designed to select between the results of SSD and CSD systems based on an analysis of the instability of the SSD system performances. Experiments on the conversational telephone speech (CTS) data from DIHARD-III Challenge show that the proposed SGSD system can significantly improve the performance of state-of-the-art CSD systems, yielding relative diarization error rate reductions of 20.2 $\%$ and 20.8 $\%$ on the development set and evaluation set, respectively.
\end{abstract}

\section{Introduction}
Speech diarization is a task to label partitioning speech segments with classes corresponding to speaker identities, namely ``who spoke when" \cite{2021review}. It is an important front-end for many applications, such as meeting summary,  telephone conversations analysis  and  speaker based indexing \cite{diarization_application1, diarization_application3}. Therefore, there are many different domains to evaluate the diarization performances, such as broadcast news recordings \cite{broadcast}, conversational telephone speech (CTS) \cite{CTS} and meeting conversations \cite{meeting}.

Conventional clustering-based methods are widely used in speaker diarization \cite{CD1,CD2}. The core of this technique is extracting and clustering speaker representations which mainly include  i-vector \cite{ivector} and some neural network based embeddings, such as d-vector \cite{dvector} and x-vector \cite{xvector}. In these algorithms, variational Bayesian hidden Markov model with x-vectors (VBx) \cite{BUT1} has achieved a superior performance and ranked first in DIHARD-II Challenge \cite{dihard2}. However, diarization based on speaker clustering cannot well handle overlapping speech because a speech segment can only be assigned to one specific cluster. To solve this problem, end-to-end neural speaker diarization (EEND)  \cite{EEND,EENDEDA,EENDSC} approaches were proposed where the diarization results can be directly predicted by neural networks, which allows the system to be optimized by minimizing diarization errors. Target-speaker voice activity detection (TS-VAD) \cite{TSVAD} is then proposed to predict an activity of each speaker at each time frame using speech features along with speaker embedding. All these techniques are capable of dealing with overlapping speech based on classification networks.

Speech separation (SS) is a task of separating target speech from background interferences \cite{sep1}. Most speech separation approaches have been formulated in the time-frequency (T-F) domain. Such models include deep neural networks (DNNs) \cite{sepdnn}, recurrent neural networks (RNNs) \cite{seprnn,seprnn1}, and generative adversarial networks (GANs) \cite{sepgan}. Recently,  time-domain based networks, such as fully-convolutional time-domain audio separation network (Conv-TasNet) \cite{Tasnet} and dual-path RNN \cite{Dualpathrnn}, have shown good results in speech separation. However, most of above algorithms were evaluated on simulation data. For data sets with realistic conditions, the separation performances cannot be directly measured due to the lack of clean speech. This limits the application of separation to diarization tasks based on realistic data sets.

In separation-based speaker diarization (SSD), we can separate the utterances by trained Conv-TasNets and obtain diarization results by detecting the speech segments in separated speech.  However, when dealing with realistic mismatched data, the separation performances are often unstable, which leads to worse diarization performances than those obtained with clustering-based speaker diarization (CSD) systems. Therefore, we analyze different cases in SSD systems and propose some strategies, including speech duration checking, overlap ratio checking and relative diarization error rate (DER) calculation. We call this improvement separation guided speaker diarization (SGSD) approach which enables separation to assist CSD in handling overlap regions. Through these strategies, SGSD can indirectly measure the separation performances in realistic utterances and select between the results of SSD and CSD.

Experiments on CTS dataset show that the proposed SGSD system can help CSD achieve a good performance on overlap regions. Similar works exist in \cite{sepmeeting1, sepmeeting2}. However, our proposed SGSD framework offers a few major differences: (1) different from the works in \cite{sepmeeting1,sepmeeting2} which use the BLSTM based separation model, we adopt more powerful Conv-TasNet separation model. It avoids the assumption that the speaker masks are additive and sum to one for each time-frequency bin which is not directly applicable to diarization \cite{EENDSC}; (2) we evaluate our methods on realistic mismatched single-channel dataset with different speaking styles from our training set, which is more challenging than handling the simulated single-channel data in \cite{sepmeeting1} and multi-channel dataset with similar speaking styles to training set in \cite{sepmeeting2}; and (3) due to the more challenging situation, we cannot directly use speech separation to attain the diarization results. Therefore, different from the multi-task perspective in \cite{sepmeeting1,sepmeeting2}, we emphasize the aspect of enabling speech separation to assist CSD in the proposed SGSD system.
\begin{figure*}[t]
	\centering
	\includegraphics[width=\linewidth]{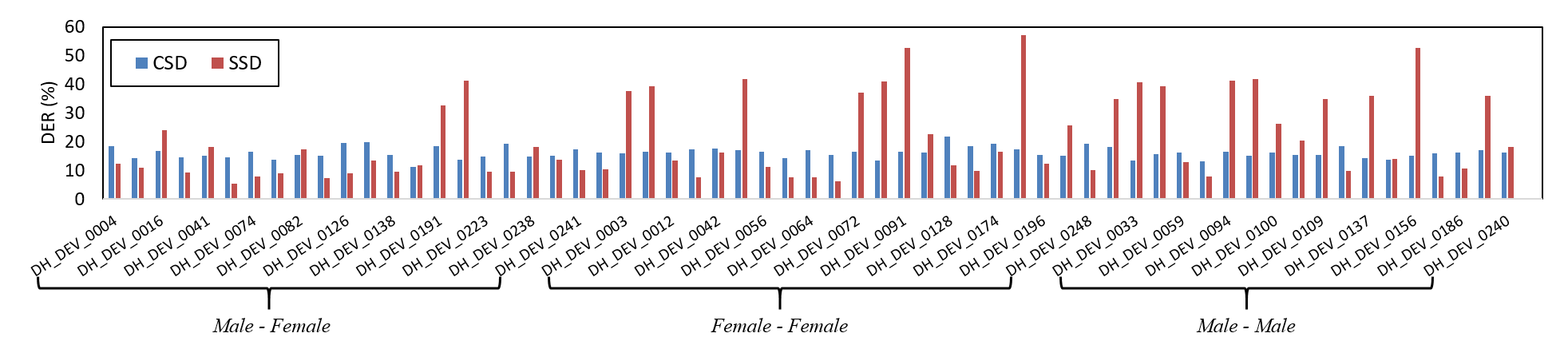}
	\vspace{-0.70cm}
	\caption{Performance (DER) comparison between CSD system and SSD system on CTS domain of DIHARD-III development set.}
	\label{result_compare_all}
\end{figure*}
\section{The Proposed SGSD Framework}
\begin{figure}[htbp]
	\centering
	\includegraphics[width=1.05\linewidth]{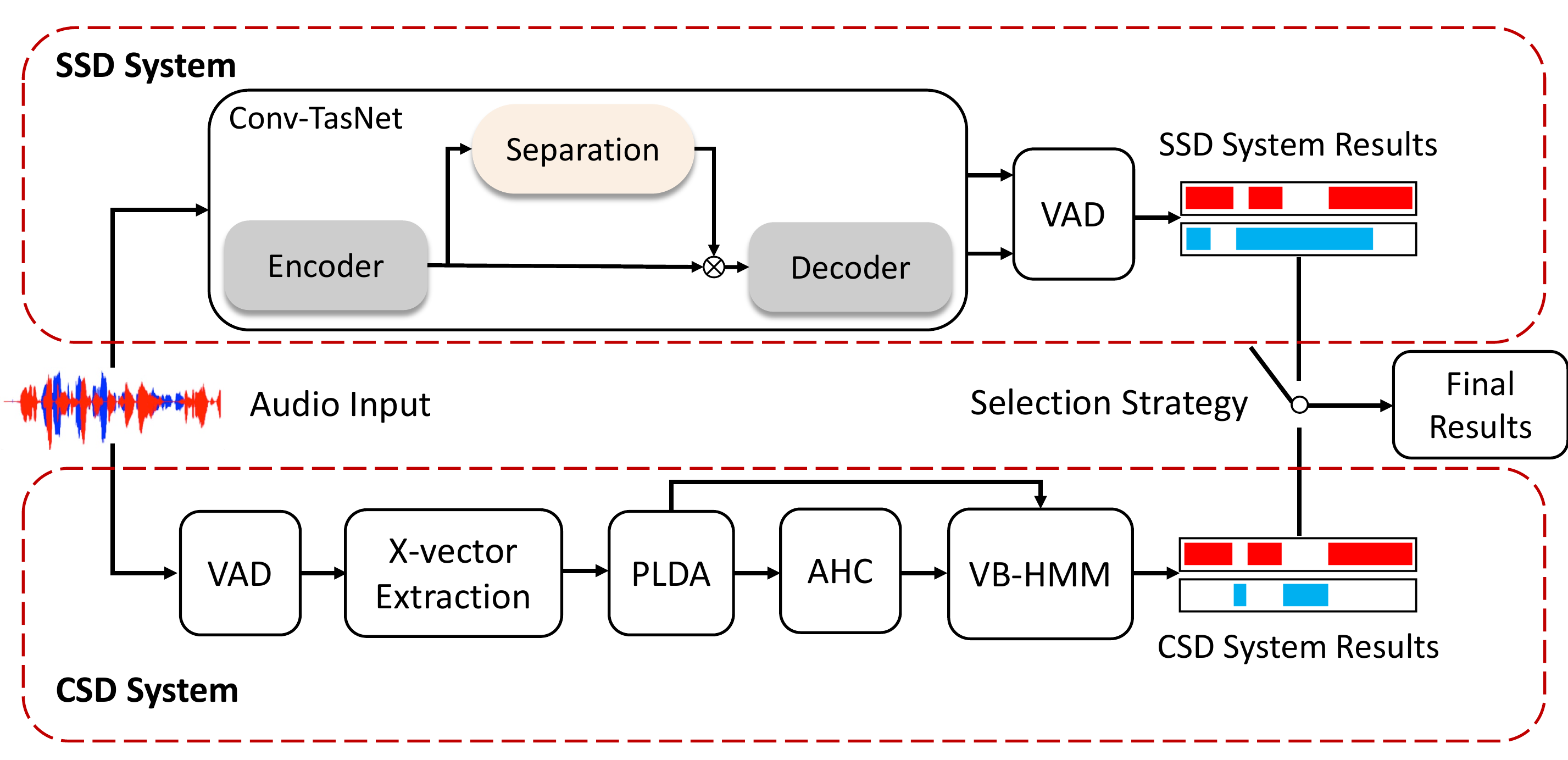}
	\vspace{-0.5cm}
	\caption{Flowchart of separation guided speaker diarization}
	\label{whole frame}
\end{figure}

Fig. \ref{whole frame} shows a flowchart of the proposed SGSD system. As can be seen, it contains two single diarization systems: clustering-based speaker diarization (CSD) and separation-based speaker diarization (SSD).  We design several strategies to better select between the results of these two systems, which enables the SSD to assist CSD in dealing with overlap regions. Details of the proposed system are described in the following subsections.

\vspace{-0.5cm}

\subsection{Clustering-based speaker diarization (CSD)}
As shown in the bottom of Fig.\ref{whole frame}, we use VBx \cite{BUT1} as our clustering-based speaker diarization (CSD) system. Apart from the conventional processes which include voice activity detection (VAD), speaker feature extraction and speaker clustering, the VBx also employs  VB-HMM to refine the assignments of x-vectors to speaker clusters.

\subsection{Separation-based speaker diarization (SSD)}
The top of Fig. \ref{whole frame} illustrates the framework of the separation-based speaker diarization (SSD) system. As can be seen in the figure, the SSD system simply contains two parts: separation and detection.
In separation part, original utterance is separated into two streams by Conv-TasNet \cite{Tasnet} based separation model. Ideally, overlapping speech segments are automatically separated, and the single-speaker speech segments are assigned to the streams corresponding to speaker identities. In this part, the utterance-level permutation invariant training (uPIT) based learning objective is used to optimize the model parameters:
\begin{equation}\label{eq0}
L = \frac{1}{N}\sum_{i=1}^{N}l(\hat{s}_i - s_{\phi})
\end{equation}
where $N$ is the number of speakers which is 2 in this study, $l$ is the error between the network output and the target, $\hat{s}_i$ denotes $i$-th predicted speech, $s_{\phi}$ denotes reference speech with the permutation $\phi$ that minimizes the training objective $L$. In training, $l$ is calculated by scale-invariant source-to-noise ratio (Si-SNR), which is an evaluation metric for source separation:
\begin{equation}\label{eq1}
		\rm{Si}\emph{-}{SNR} = 10\log_{10}\frac{||\rm{s}_{\text{target}}||^2}{||\hat{s} - \rm{s}_{\text{target}}||^2}
\end{equation}
where $\rm{s}_{\text{target}} =  \frac{<\hat{s}, s>s}{||s||^2}$. $\rm{\hat{s}}$ and $\rm{s}$ are the estimates and targets respectively.
In detection part, the separated two-stream speech signals are sent to the VAD model to get the time label of speech segments. Combine all detection results along the time axis, then speaker diarization results are obtained.

\subsection{Separation guided speaker diarization (SGSD)}
Fig.\ref{result_compare_all} presents the performance (DER) comparison between CSD and SSD systems on each utterance from CTS domain of DIHARD-III development set.  It can be observed that the performance of SSD system is unstable. On about half of the utterances, the SSD results are better than CSD results, while in other utterances, the SSD performance is fairly poor, with DERs even greater than 50$\%$. On the contrary, the CSD system is relatively stable in all utterances. Therefore, we can see that both systems have advantages and disadvantages: SSD system can handle the overlap regions while its stability is poor, the CSD system is pretty stable while it cannot well process the overlapping speech. The motivation of our method is utilizing the complementarity of these two systems. What's more, from the Fig. \ref{whole frame} we can see that poor separation performance will lead to the degradation of SSD performance. However, when dealing with the realistic mismatched datasets, the speech separation performance is unstable, and the separation performance cannot be directly measured in the realistic datasets due to the lack of clean speech.
Therefore, we need some strategies which can indirectly measure the separation performance to better select between the results of the CSD and SSD systems.
Inspired by these facts, the proposed SGSD procedure is illustrated in Algorithm 1.
First, we get the diarization results of SSD and CSD systems for all utterances. 
\vspace{-0.2cm}
\begin{algorithm}[h]
	\caption{SGSD Procedure}
	\vspace{0.02in}
	\hspace*{0.02in} {\bf Step1: Results generation}\\
	\vspace{0.02in}
	\hspace*{0.02in} Obtain the SSD and CSD results for all utterances;\\
	\hspace*{0.02in} {\bf Step2: Performance measuring} \\
	\vspace{0.02in}
	\hspace*{0.02in} Use the SSD results to measure the SS performance;\\
	\hspace*{0.02in} {\bf Step3: Utterances capture} \\
	\vspace{0.02in}
	\hspace*{0.02in} Capture the utterances with poor SS performance; \\
	\hspace*{0.02in} {\bf Step4: Results selection} \\
	\vspace{0.02in}
	\hspace*{0.02in} Use the CSD results for the selected utterances;\\
	\vspace{0.0in}
	\hspace*{0.02in} Use the SSD results for the other utterances.
\end{algorithm}
For some utterances, the separation part of SSD system doesn't output the correct single-speaker speech. Next, we use the SSD results to measure the performance of speech separation. Then, we select the utterances which are judged to have poor speech separation results. Finally, we adopt the CSD results for selected utterances and the SSD results for unselected utterances. In SGSD,  we use the diarization results to indirectly measure the separation performance in SSD system because poor separation results tend to lead to unreasonable diarization results.
By analyzing some failure cases in separation part of SSD system (this will be described in detail in Section 3), we propose some strategies to measure the separation performance in Step 2.

To better illustrate the strategies, we assume that $N$ denotes the number of speakers, $R_i$ denotes the regions of speaker $i$ and $\emph{d}(R)$ denotes the duration of the region $R$. Specially, $N=2$ in the CTS dataset.  We propose three strategies which will be introduced one by one.

\vspace{0.1cm}
\noindent\textbf{Strategy 1}: Check if the duration of speakers' speech is unbalanced in SSD results:
\begin{equation}\label{eq2}
	\frac{\rm{min}(\emph{d}(\emph{R}_\emph{1}), \emph{d}(\emph{R}_\emph{2}), ..., \emph{d}(\emph{R}_\emph{N}))}{\rm{max}(\emph{d}(\emph{R}_\emph{1}), \emph{d}(\emph{R}_\emph{2}), ..., \emph{d}(\emph{R}_\emph{N}))} > th_1
\end{equation}
the $th_1$ is the threshold which can be adjusted. If the InEq. (\ref{eq2}) is unsatisfied, it means that the duration of the two speakers is unbalanced, and the performance of speech separation will be judged as poor.

\vspace{0.15cm}
\noindent\textbf{Strategy 2}: Check if there is an abnormal overlap ratio in SSD results:
\begin{equation}\label{eq3}
	\frac{\sum_{\emph{i}=1}^{N}\emph{d}(\emph{R}_\emph{i}) -  \emph{d}(\emph{R}_\emph{1} \cup \emph{R}_\emph{2} \cup \cdots \cup \emph{R}_\emph{N} )}{\sum_{\emph{i}=1}^{N}\emph{d}(\emph{R}_\emph{i})} < th_2
\end{equation}
The left side of the InEq. (\ref{eq3}) is the overlap ratio of an utterance \cite{OLR}.
If the InEq. (\ref{eq3}) is unsatisfied when $th_2$ is set to an appropriate value, it implies that the overlap ratio in the SSD results is too large, and the separation results will be judged as incorrect.

\vspace{0.15cm}
\noindent\textbf{Strategy 3}: Calculate the deviation degree of the SSD results relative to the CSD results, namely:
\begin{equation}\label{eq4}
	\frac{\sum_{s=1}^{S}\emph{d}(s)\cdot(\rm{max}(\emph{K}_{\emph{CSD}}(\emph{s}),\emph{K}_{\emph{SSD}}(\emph{s}))- \emph{K}(\emph{s}))}{\sum_{s=1}^{S}\emph{d}(\emph{s}   )\cdot\emph{K}_{\emph{CSD}}(s)} < th_3
\end{equation}
where $S$ is the number of speaker segments in which both CSD results and SSD results contain the same speaker (or speakers). $\emph{K}_{\emph{CSD}}(\emph{s})$ and $\emph{K}_{\emph{SSD}}(\emph{s})$ denote the speaker number in speech segment $s$ of CSD and SSD results respectively. $\emph{K}(\emph{s})$ means the number of speakers in speech segment $s$ that are correctly matched between CSD and SSD results. This is actually the calculation of DER \cite{DER}, but we replace the ground truth with the CSD results.  If the InEq. (\ref{eq4}) is unsatisfied, it means the SSD results greatly deviate from the stable CSD results and the corresponding separation performance is poor.
\vspace{0.1cm}

Among them, SGSD with the first strategy (hereinafter referred to as SGSD1) and the second strategy (hereinafter referred to as SGSD2) measure the separation performance based on the diarization results of the SSD system itself, and SGSD with the third strategy (hereinafter referred to as SGSD3) uses the clustering-based method as a benchmark to measure the separation performance of SSD system. Through the above three strategies, we can detect the utterances with poor separation performance in SSD system.

\section{Experiments}
\subsection{Experimental conditions}
The training set of the separation model in SSD system was simulated on Librispeech \cite{librispeech} dataset. We randomly selected two utterances from different speakers in Librispeech dataset and mixed them to obtain the simulated training utterance. In this paper, we simulated about 250-hour training data.
To verify the effectiveness of the proposed method, we conducted the experiments on the realistic conversational telephone speech (CTS) dataset from development set and evaluation set of DIHARD-III Challenge \cite{dihard3}. Both sets contain 61 utterances with each utterance consisting of a 10-minute conversation between two native English speakers.
The overlap ratio of the development set and evaluation set is 11.9$\%$ and 10.5$\%$ respectively, which is quite large in the two-speaker domain.
We compared the proposed method with VBx system \cite{but} and referred to the configuration used in
recipe\footnote{https://github.com/BUTSpeechFIT/VBx/tree/v1.0\underline{~~}DIHARDII} published by BUT speech team.
In the separation part of SSD system,  we used the asteroid toolkit \cite{asteroid} to train a Conv-TasNet as separation model. We trained the model for 75 epochs on 3-second long segments. The learning rate was set to 0.001, and the batch-size was set to 6. Adam \cite{adam} was adopted as the optimizer. Moreover, WebRTC VAD\footnote{https://github.com/wiseman/py-webrtcvad} with 30ms hop length was employed in the detection part. In SGSD1, we set the minimum ratio of the duration of two speakers ($th_1$) to 40 $\%$. In SGSD2, we set the highest overlap ratio ($th_2$) in SSD results to 20 $\%$. In SGSD3, we set the maximum relative deviation ($th_3$) to 26 $\%$. These thresholds were determined based on the development set. Diarization error rate (DER) \cite{DER} was used as evaluation metric in our experiments. It consists of computing speaker error, false alarm and missed speech. We included all errors in calculation of DER. In addition, we didn't set any forgiveness collar during evaluation.

\begin{figure*}[t]
	\centering
	\subfigure[\emph{Successful case  (DER = 4.6 $\%$)}]{
		\begin{minipage}[t]{0.33\linewidth}
			\centering
			\includegraphics[width=\linewidth]{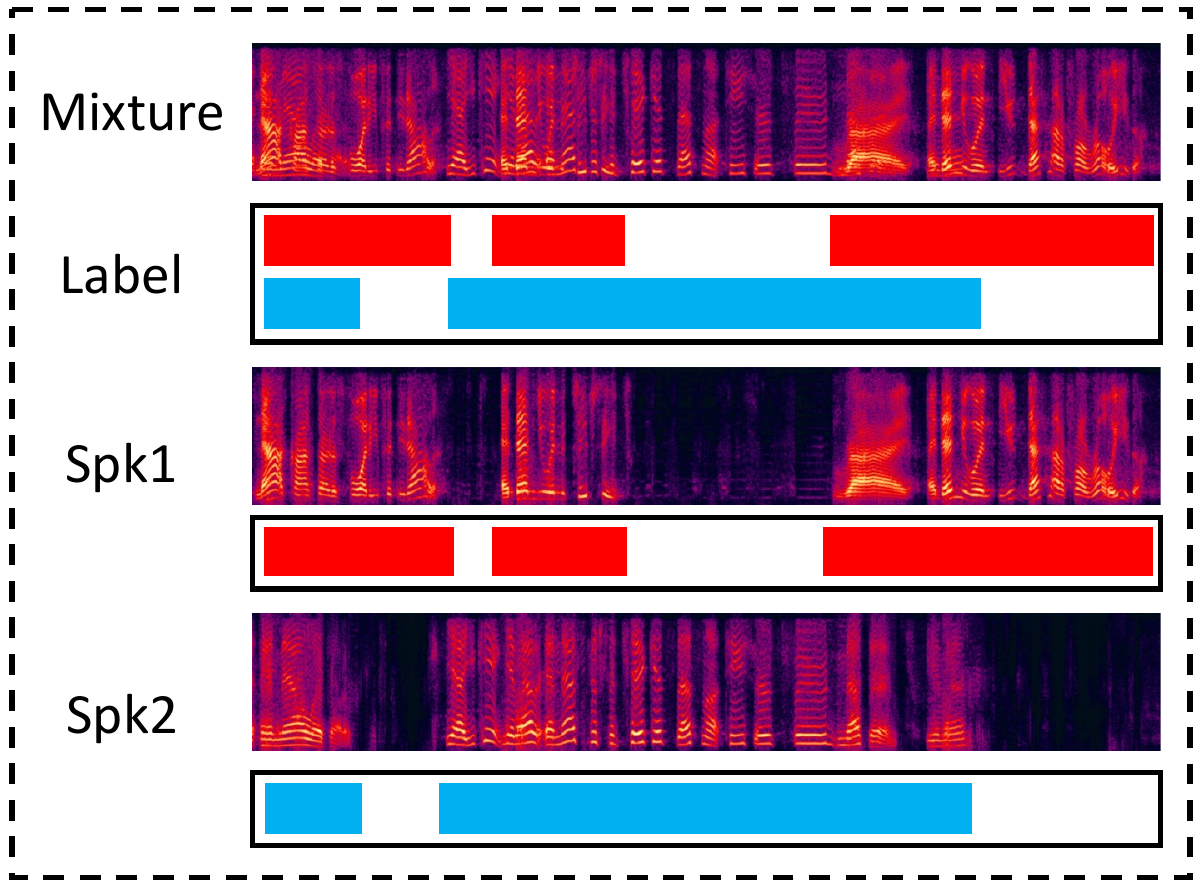}
		\end{minipage}%
	}%
	\subfigure[\emph{Failure case 1 (DER = 42.5 $\%$)}]{
		\begin{minipage}[t]{0.33\linewidth}
			\centering
			\includegraphics[width=\linewidth]{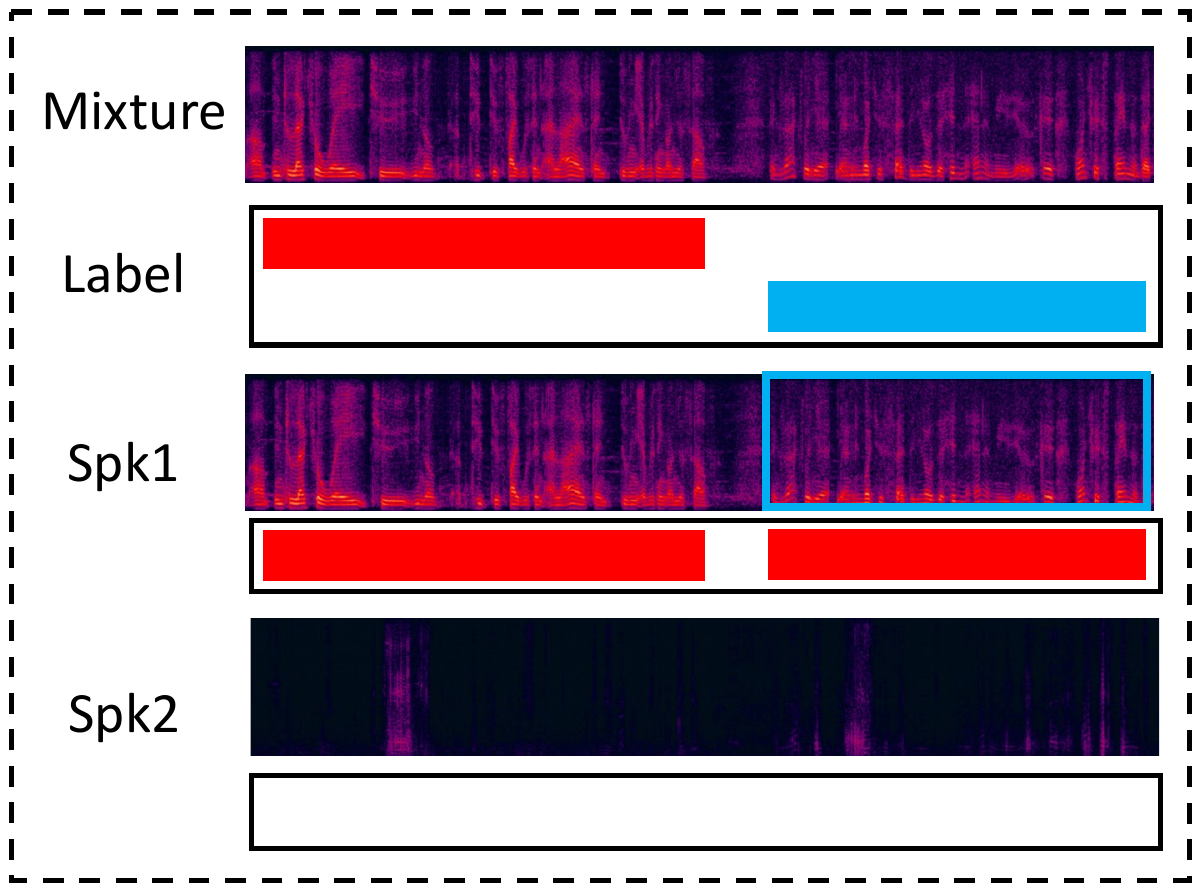}
		\end{minipage}%
	}%
	\subfigure[\emph{Failure case 2 (DER = 82.4 $\%$)}]{
		\begin{minipage}[t]{0.33\linewidth}
			\centering
			\includegraphics[width=\linewidth]{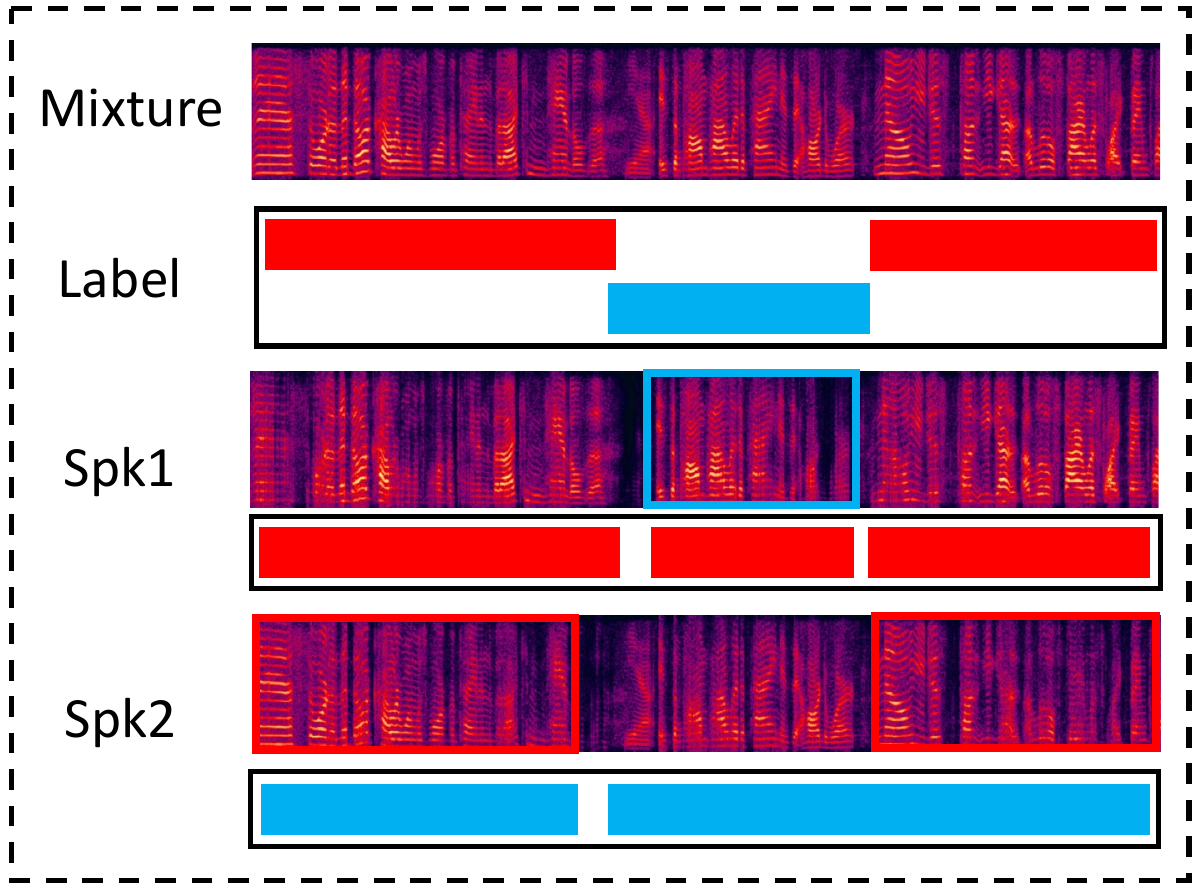}
		\end{minipage}%
	}%
	\vspace{-0.2cm}
	\caption{Spectrograms and diarization labels of speech segments from three selected utterances in SSD system (the rectangles mark the regions which were falsely separated) . Case (a): SS successfully separates a two-speaker mixed utterance, Cases (b) and (c): two SS failures.}
	\label{different case}
\end{figure*}

\subsection{Analysis of SSD system under mismatched conditions}
Although we used the powerful Conv-TasNet based separation model in SSD system, the separation performance was still unstable due to the mismatch between simulated reading style training set and realistic conversational style test set. In order to propose suitable selection strategies, we analyzed the different cases of separation results in SSD system. Fig. \ref{different case} presents the spectrograms and diarization labels of 10s speech segments from three selected utterances which belong to different cases.
Fig. \ref{different case} (a) shows a successful case where the separation result has a clear relationship with the diarization label. In this case, the DER is quite small (DER = 4.6 $\%$).  Fig. \ref{different case} (b) shows the failure case where the speech segments of different speakers are assigned to the same stream, which leads to a large speaker error (SpkErr = 36.1 $\%$). Fig. \ref{different case} (c) shows the failure case where the speech segments of one speaker are assigned to two streams, which leads to a large false alarm error (FA = 78.1 $\%$). From these cases we can see, in SSD system, successful separation will yield the good diarization results while failure speech separation will lead to a large speaker error (failure case 1) or false alarm error (failure case 2). In addition, we found that there was a strong relationship between the speaker gender combination and diarization performance in SSD system as shown in the bottom of Fig. \ref{result_compare_all}.  For utterances with different gender combination (Male-Female), the SSD results are often better than CSD results. Conversely, for utterances with the same gender combination (Female-Female or Male-Male), the performance of SSD system is often very poor. As we know, the same gender speaker mix is more difficult to separate than the case of different gender mix in speech separation \cite{gender}. We can observe the consistency between SSD and SS performance, which also verifies that poor separation results will cause the degradation of SSD system performance as mentioned in Section 2.3.

Moreover, from Fig. \ref{different case} we can see that if the separation result belongs to the first failure case, the duration of the two speakers will be very unbalanced (e.g., the ratio of two speakers' duration is 6.3$\%$ in SSD result for the segment shown in Fig. \ref{different case} (b)). This corresponds to Strategy 1 (SGSD1) in Section 2.3. If the separation result belongs to the second failure case, the overlap ratio of the SSD result will be too high (e.g., the overlap ratio is 84.2$\%$ in SSD for the segment shown in Fig. \ref{different case} (c)), which corresponds to Strategy 2 (SGSD2) in Section 2.3.

\subsection{Overall comparison}
Table \ref{tab:measure} compares the performance of detecting the utterances with poor separation performance (i.e., SSD results are worse than CSD results) among different SGSD systems on the CTS development and evaluation sets from DIHARD-III Challenge. ``SGSD1$\&$2" means combining the detection results of SGSD1 and SGSD2. ``SGSD1$\&$2$\&$3'' means  voting on SGSD1, SGSD2 and SGSD3.
From this table we can make several observations. First, by comparing the different SGSD systems,  SGSD3 has achieved the best performance on both development set and evaluation set, which indicates that DER between SSD and CSD is a good and robust indicator for measuring the speech separation performance.
Second, combining the results of SGSD1 and SGSD2 can significantly improve the detection performance, which illustrates the complementarity of them. Third, compared with the SGSD3, the voting of SGSD1, SGSD2 and SGSD3 leads to the worse detection performance due to the detection errors of SGSD1 and SGSD2. However, even with our best system SGSD3, some utterances with relatively poor speech separation performance are not detected because the differences between CSD results and SSD results in these utterances are small. Generally speaking, these SSD results are not too poor due to the small differences from stable CSD results.
Table \ref{tab:final_result} compares the DERs of CSD, SGSD3 and Oracle (perfect detection / selection between SSD and CSD with ground truth). It can be observed that the speaker errors of CSD results are quite small in both development set and evaluation set (4.2 $\%$ and 3.7 $\%$ respectively), which means CSD system can achieve good performance on CTS dataset. However, due to the lack of ability to handle the overlapping speech, most of the DER comes from the miss error. What's more, SGSD3 can help CSD system handle the overlapping speech which can be seen from the smaller miss error of SGSD3 results compared with CSD results. It is worth noting that the SGSD3 has achieved quite good results which is very close to the oracle results.

\begin{table}[t]
	\renewcommand\arraystretch{1.25}
	\centering
	\caption{Detection comparison among different SGSD systems on the CTS domain of development set and evaluation set from DIHARD-III Challenge. }
	\vspace{-0.3cm}
	\label{tab:measure}\medskip
	\resizebox{8.0cm}{!}{\begin{tabular}{c|c|c|c|c|c|c}
			\hline
			\multirow{2}*{Method}&\multicolumn{3}{c}{Dev}&\multicolumn{3}{|c}{Eval}\\
			\cline{2-7}
			~&Recall&Precision&Acc&Recall&Precision&Acc\\
			\hline
			SGSD1&0.35&1.0&0.69&0.43&0.92&0.72\\
			\hline
			SGSD2&0.55&1.0&0.79&0.36&1.0&0.71\\
			\hline
			SGSD3&0.90&0.93&\textbf{0.92}&0.86&0.92&\textbf{0.90}\\
			\hline
			SGSD1$\&$2&0.79&1.0&0.90&0.71&0.95&0.85\\
			\hline
			SGSD1$\&$2$\&$3&0.79&1.0&0.90&0.71&1.0&0.87\\
			\hline
	\end{tabular}}
\end{table}

\begin{table}[t]
	\renewcommand\arraystretch{1.25}
	\centering
	\caption{Performance comparison among CSD system, SGSD system with Strategy 3 (denoted as SGSD3) and oracle system (denoted as Oracle).}
	\vspace{-0.3cm}
	\label{tab:final_result}\medskip
	\resizebox{7.7cm}{!}{\begin{tabular}{c|c|c|c|c|c}
			\hline
			Set&Method&MISS ($\%$)&FA ($\%$)&SpkErr ($\%$)&DER ($\%$)\\
			\hline
			\multirow{3}*{Dev}&CSD&12.0&0.0&4.2&16.22\\
			\cline{2-6}
			~&SGSD3&7.6&2.6&2.7&12.95\\
			\cline{2-6}
			~&Oracle&7.5&2.6&2.6&12.75\\
			\hline
			\multirow{3}*{Eval}&CSD&10.5&0.0&3.7&14.20\\
			\cline{2-6}
			~&SGSD3&6.4&2.6&2.2&11.24\\
			\cline{2-6}
			~&Oracle&6.4&2.4&2.2&10.94\\
			\hline
	\end{tabular}}
\end{table}

\section{Conclusion}
We propose a SGSD approach to enabling separation-based processing to assist clustering-based systems in handling overlapping speech regions. To reduce the impact of the instability of separation performance, we design some strategies to select between the results of CSD and SSD systems. Experiments on the CTS data show that the proposed SGSD can help improve the conventional clustering-based systems. In the future, we will explore SGSD approaches under more challenging multi-speaker (more than 2 speakers) and noisy conditions.

\pagebreak
\bibliographystyle{IEEEtran}

\bibliography{mybib}

\begin{thebibliography}{10}
\providecommand{\url}[1]{#1}
\csname url@samestyle\endcsname
\providecommand{\newblock}{\relax}
\providecommand{\bibinfo}[2]{#2}
\providecommand{\BIBentrySTDinterwordspacing}{\spaceskip=0pt\relax}
\providecommand{\BIBentryALTinterwordstretchfactor}{4}
\providecommand{\BIBentryALTinterwordspacing}{\spaceskip=\fontdimen2\font plus
\BIBentryALTinterwordstretchfactor\fontdimen3\font minus
  \fontdimen4\font\relax}
\providecommand{\BIBforeignlanguage}[2]{{%
\expandafter\ifx\csname l@#1\endcsname\relax
\typeout{** WARNING: IEEEtran.bst: No hyphenation pattern has been}%
\typeout{** loaded for the language `#1'. Using the pattern for}%
\typeout{** the default language instead.}%
\else
\language=\csname l@#1\endcsname
\fi
#2}}
\providecommand{\BIBdecl}{\relax}
\BIBdecl

\bibitem{2021review}
T.~J. Park, N.~Kanda, D.~Dimitriadis, K.~J. Han, S.~Watanabe, and S.~Narayanan,
  ``A review of speaker diarization: Recent advances with deep learning,''
  \emph{arXiv preprint arXiv:2101.09624}, 2021.

\bibitem{diarization_application1}
D.~Vijayasenan, F.~Valente, and H.~Bourlard, ``An information theoretic
  approach to speaker diarization of meeting data,'' \emph{IEEE Transactions on
  Audio, Speech, and Language Processing}, vol.~17, no.~7, pp. 1382--1393,
  2009.

\bibitem{diarization_application3}
X.~Anguera, S.~Bozonnet, N.~Evans, C.~Fredouille, G.~Friedland, and O.~Vinyals,
  ``Speaker diarization: A review of recent research,'' \emph{IEEE Transactions
  on Audio, Speech, and Language Processing}, vol.~20, no.~2, pp. 356--370,
  2012.

\bibitem{broadcast}
M.~A. Siegler, U.~Jain, B.~Raj, and R.~M. Stern, ``Automatic segmentation,
  classification and clustering of broadcast news audio,'' in \emph{Proc. DARPA
  speech recognition workshop}, vol. 1997, 1997.

\bibitem{CTS}
P.~Kenny, D.~Reynolds, and F.~Castaldo, ``Diarization of telephone
  conversations using factor analysis,'' \emph{IEEE Journal of Selected Topics
  in Signal Processing}, vol.~4, no.~6, pp. 1059--1070, 2010.

\bibitem{meeting}
D.~Vijayasenan, F.~Valente, and H.~Bourlard, ``An information theoretic
  approach to speaker diarization of meeting data,'' \emph{IEEE Transactions on
  Audio, Speech, and Language Processing}, vol.~17, no.~7, pp. 1382--1393,
  2009.

\bibitem{CD1}
S.~H. Shum, N.~Dehak, R.~Dehak, and J.~R. Glass, ``Unsupervised methods for
  speaker diarization: An integrated and iterative approach,'' \emph{IEEE
  Transactions on Audio, Speech, and Language Processing}, vol.~21, no.~10, pp.
  2015--2028, 2013.

\bibitem{CD2}
G.~Sell and D.~Garcia-Romero, ``Speaker diarization with plda i-vector scoring
  and unsupervised calibration,'' in \emph{2014 IEEE Spoken Language Technology
  Workshop (SLT)}.\hskip 1em plus 0.5em minus 0.4em\relax IEEE, 2014, pp.
  413--417.

\bibitem{ivector}
N.~Dehak, P.~J. Kenny, R.~Dehak, P.~Dumouchel, and P.~Ouellet, ``Front-end
  factor analysis for speaker verification,'' \emph{IEEE Transactions on Audio,
  Speech, and Language Processing}, vol.~19, no.~4, pp. 788--798, 2010.

\bibitem{dvector}
L.~Wan, Q.~Wang, A.~Papir, and I.~L. Moreno, ``Generalized end-to-end loss for
  speaker verification,'' in \emph{2018 IEEE International Conference on
  Acoustics, Speech and Signal Processing (ICASSP)}.\hskip 1em plus 0.5em minus
  0.4em\relax IEEE, 2018, pp. 4879--4883.

\bibitem{xvector}
D.~Snyder, D.~Garcia-Romero, G.~Sell, D.~Povey, and S.~Khudanpur, ``X-vectors:
  Robust dnn embeddings for speaker recognition,'' in \emph{2018 IEEE
  International Conference on Acoustics, Speech and Signal Processing
  (ICASSP)}.\hskip 1em plus 0.5em minus 0.4em\relax IEEE, 2018, pp. 5329--5333.

\bibitem{BUT1}
M.~Diez, L.~Burget, F.~Landini, and J.~{\v{C}}ernock{\`y}, ``Analysis of
  speaker diarization based on bayesian hmm with eigenvoice priors,''
  \emph{IEEE/ACM Transactions on Audio, Speech, and Language Processing},
  vol.~28, pp. 355--368, 2019.

\bibitem{dihard2}
N.~Ryant, K.~Church, C.~Cieri, A.~Cristia, J.~Du, S.~Ganapathy, and
  M.~Liberman, ``Second dihard challenge evaluation plan,'' \emph{Linguistic
  Data Consortium, Tech. Rep}, 2019.

\bibitem{EEND}
Y.~Fujita, N.~Kanda, S.~Horiguchi, K.~Nagamatsu, and S.~Watanabe, ``End-to-end
  neural speaker diarization with permutation-free objectives,'' \emph{arXiv
  preprint arXiv:1909.05952}, 2019.

\bibitem{EENDEDA}
S.~Horiguchi, Y.~Fujita, S.~Watanabe, Y.~Xue, and K.~Nagamatsu, ``End-to-end
  speaker diarization for an unknown number of speakers with encoder-decoder
  based attractors,'' \emph{arXiv preprint arXiv:2005.09921}, 2020.

\bibitem{EENDSC}
Y.~Fujita, S.~Watanabe, S.~Horiguchi, Y.~Xue, J.~Shi, and K.~Nagamatsu,
  ``Neural speaker diarization with speaker-wise chain rule,'' \emph{arXiv
  preprint arXiv:2006.01796}, 2020.

\bibitem{TSVAD}
I.~Medennikov, M.~Korenevsky, T.~Prisyach, Y.~Khokhlov, M.~Korenevskaya,
  I.~Sorokin, T.~Timofeeva, A.~Mitrofanov, A.~Andrusenko, I.~Podluzhny
  \emph{et~al.}, ``Target-speaker voice activity detection: a novel approach
  for multi-speaker diarization in a dinner party scenario,'' \emph{arXiv
  preprint arXiv:2005.07272}, 2020.

\bibitem{sep1}
D.~Wang and J.~Chen, ``Supervised speech separation based on deep learning: An
  overview,'' \emph{IEEE/ACM Transactions on Audio, Speech, and Language
  Processing}, vol.~26, no.~10, pp. 1702--1726, 2018.

\bibitem{sepdnn}
J.~Du, Y.~Tu, L.-R. Dai, and C.-H. Lee, ``A regression approach to
  single-channel speech separation via high-resolution deep neural networks,''
  \emph{IEEE/ACM Transactions on Audio, Speech, and Language Processing},
  vol.~24, no.~8, pp. 1424--1437, 2016.

\bibitem{seprnn}
M.~Kolb{\ae}k, D.~Yu, Z.-H. Tan, and J.~Jensen, ``Multitalker speech separation
  with utterance-level permutation invariant training of deep recurrent neural
  networks,'' \emph{IEEE/ACM Transactions on Audio, Speech, and Language
  Processing}, vol.~25, no.~10, pp. 1901--1913, 2017.

\bibitem{seprnn1}
Y.~Isik, J.~L. Roux, Z.~Chen, S.~Watanabe, and J.~R. Hershey, ``Single-channel
  multi-speaker separation using deep clustering,'' \emph{arXiv preprint
  arXiv:1607.02173}, 2016.

\bibitem{sepgan}
C.~Li, L.~Zhu, S.~Xu, P.~Gao, and B.~Xu, ``Cbldnn-based speaker-independent
  speech separation via generative adversarial training,'' in \emph{2018 IEEE
  International Conference on Acoustics, Speech and Signal Processing
  (ICASSP)}.\hskip 1em plus 0.5em minus 0.4em\relax IEEE, 2018, pp. 711--715.

\bibitem{Tasnet}
Y.~Luo and N.~Mesgarani, ``Conv-tasnet: Surpassing ideal time--frequency
  magnitude masking for speech separation,'' \emph{IEEE/ACM transactions on
  audio, speech, and language processing}, vol.~27, no.~8, pp. 1256--1266,
  2019.

\bibitem{Dualpathrnn}
Y.~Luo, Z.~Chen, and T.~Yoshioka, ``Dual-path rnn: efficient long sequence
  modeling for time-domain single-channel speech separation,'' in \emph{ICASSP
  2020-2020 IEEE International Conference on Acoustics, Speech and Signal
  Processing (ICASSP)}.\hskip 1em plus 0.5em minus 0.4em\relax IEEE, 2020, pp.
  46--50.

\bibitem{sepmeeting1}
T.~von Neumann, K.~Kinoshita, M.~Delcroix, S.~Araki, T.~Nakatani, and
  R.~Haeb-Umbach, ``All-neural online source separation, counting, and
  diarization for meeting analysis,'' in \emph{ICASSP 2019-2019 IEEE
  International Conference on Acoustics, Speech and Signal Processing
  (ICASSP)}.\hskip 1em plus 0.5em minus 0.4em\relax IEEE, 2019, pp. 91--95.

\bibitem{sepmeeting2}
K.~Kinoshita, M.~Delcroix, S.~Araki, and T.~Nakatani, ``Tackling real noisy
  reverberant meetings with all-neural source separation, counting, and
  diarization system,'' in \emph{ICASSP 2020-2020 IEEE International Conference
  on Acoustics, Speech and Signal Processing (ICASSP)}.\hskip 1em plus 0.5em
  minus 0.4em\relax IEEE, 2020, pp. 381--385.

\bibitem{OLR}
Q.~Lin, W.~Cai, L.~Yang, J.~Wang, J.~Zhang, and M.~Li, ``Dihard ii is still
  hard: Experimental results and discussions from the dku-lenovo team,''
  \emph{arXiv preprint arXiv:2002.12761}, 2020.

\bibitem{DER}
``The 2009 (rt-09) rich transcription meeting recognition evaluation plan,''
  \emph{http://www.itl.nist.gov/iad/mig/tests/rt/2009/docs/rt09-meeting-eval-plan-v2.pdf},
  2009.

\bibitem{librispeech}
V.~Panayotov, G.~Chen, D.~Povey, and S.~Khudanpur, ``Librispeech: an asr corpus
  based on public domain audio books,'' in \emph{2015 IEEE international
  conference on acoustics, speech and signal processing (ICASSP)}.\hskip 1em
  plus 0.5em minus 0.4em\relax IEEE, 2015, pp. 5206--5210.

\bibitem{dihard3}
N.~Ryant, K.~Church, C.~Cieri, J.~Du, S.~Ganapathy, and M.~Liberman, ``Third
  dihard challenge evaluation plan,'' \emph{arXiv preprint arXiv:2006.05815},
  2020.

\bibitem{but}
F.~Landini, S.~Wang, M.~Diez, L.~Burget, P.~Mat{\v{e}}jka,
  K.~{\v{Z}}mol{\'\i}kov{\'a}, L.~Mo{\v{s}}ner, A.~Silnova, O.~Plchot,
  O.~Novotn{\`y} \emph{et~al.}, ``But system for the second dihard speech
  diarization challenge,'' in \emph{ICASSP 2020-2020 IEEE International
  Conference on Acoustics, Speech and Signal Processing (ICASSP)}.\hskip 1em
  plus 0.5em minus 0.4em\relax IEEE, 2020, pp. 6529--6533.

\bibitem{asteroid}
M.~Pariente, S.~Cornell, J.~Cosentino, S.~Sivasankaran, E.~Tzinis,
  J.~Heitkaemper, M.~Olvera, F.-R. St{\"o}ter, M.~Hu, J.~M.
  Mart{\'\i}n-Do{\~n}as \emph{et~al.}, ``Asteroid: the pytorch-based audio
  source separation toolkit for researchers,'' \emph{arXiv preprint
  arXiv:2005.04132}, 2020.

\bibitem{adam}
D.~P. Kingma and J.~Ba, ``Adam: A method for stochastic optimization,''
  \emph{arXiv preprint arXiv:1412.6980}, 2014.

\bibitem{gender}
K.~Wang, F.~Soong, and L.~Xie, ``A pitch-aware approach to single-channel
  speech separation,'' in \emph{ICASSP 2019-2019 IEEE International Conference
  on Acoustics, Speech and Signal Processing (ICASSP)}.\hskip 1em plus 0.5em
  minus 0.4em\relax IEEE, 2019, pp. 296--300.

\end{thebibliography}


\end{document}